\newcommand{\idty}{{\leavevmode{\rm 1\mkern -5.4mu I}}}
\newcommand{\Ir}{Z\!\!\!Z}
\newcommand{\Ibb}[1]{ {\rm I\ifmmode\mkern
            -3.6mu\else\kern -.2em\fi#1}}
\newcommand{\ibb}[1]{\leavevmode\hbox{\kern.3em\vrule
     height 1.2ex depth -.3ex width .2pt\kern-.3em\rm#1}}
\newcommand{\Cx}{{\ibb C}}
\newcommand{\Nl}{{\Ibb N}}
\newcommand{\Rl}{{\Ibb R}}
\documentstyle[12pt]{article}
\begin{document}
\renewcommand{\theequation} {\arabic{section}.\arabic{equation}}
\noindent
\centerline{\Large \bf Physical Aspects of Differential Calculi}
\vskip.4cm
\centerline{\Large \bf on Commutative Algebras$^\ast$}
\vskip1.cm
\begin{center}
      {\bf Folkert M\"uller-Hoissen}
      \vskip.1cm
      Institut f\"ur Theoretische Physik  \\
      Bunsenstr. 9, D-37073 G\"ottingen, Germany
\end{center}
\vskip.5cm
\begin{abstract}
\noindent
The central structure in various versions of noncommutative geometry
is a differential calculus on an algebra $\cal A$. This is an analogue
of the calculus of differential forms on a manifold. In this short
review we collect examples of differential calculi on {\em commutative
algebras} and explain how these are related to relevant structures in
physics.
\end{abstract}

\renewcommand{\contentsname}{\normalsize \bf Contents}
\small
\tableofcontents
\normalsize

\vskip.5cm
$\bf {}^\ast$ Lectures presented at the $XXX_q$ Karpacz Winter School for
Theoretical Physics 1994.

\section{Introduction}
\setcounter{equation}{0}
The algebra of differential forms on a manifold $\cal M$ extends the
commutative algebra of $C^\infty$-functions on a manifold to a
differential algebra. Generalizing $C^\infty ({\cal M})$ to an arbitrary
associative algebra, we may still extend it to a differential algebra
keeping the basic properties of the classical exterior derivative
(see section 2). This yields a convenient mathematical framework
to build physical models even on a `noncommutative space' in close
analogy with corresponding `classical models'.

Physical motivations to go beyond manifolds and smooth structures in
formulating physical dynamics originated in particular from ideas
about quantum gravity (see \cite{Isha89}, in particular). There
have also been suggestions towards a fundamental discreteness of
space-time (see \cite{discrete} for an incomplete list of references).
Noncommutative geometry -- and in particular algebraic differential
calculus -- appears to be the appropriate framework for pursuing
such ideas.

As an intermediate step towards differential calculus on
`noncommutative spaces' we may consider nonstandard differential
calculi on commutative spaces. This is what we will concentrate on
in these lectures (see also \cite{MH92-habil}). It will lead us to
various familiar structures in physics, but also open routes towards
new physical models.

\section{Algebraic differential calculus}
\setcounter{equation}{0}
Let $\cal A$ be an associative (and not necessarily commutative) algebra
(over $\Rl$ or $\Cx$). A {\em differential algebra} $(\Omega({\cal
A}),\mbox{d})$ over $\cal A$ is a $\Ir$-graded associative algebra
\begin{eqnarray}
   \Omega({\cal A}) = \bigoplus_{r=0}^\infty \Omega^r({\cal A})
\end{eqnarray}
where $\Omega^r({\cal A})$ are $\cal A$-bimodules and
$\Omega^0({\cal A}) = \cal A$. It is supplied with a linear operator
\begin{eqnarray}
 \mbox{d} \, : \, \Omega^r({\cal A}) \rightarrow \Omega^{r+1}({\cal A})
\end{eqnarray}
satisfying $ \mbox{d}^2=0$ and
\begin{eqnarray}             \label{Leibniz}
  \mbox{d}(\omega \omega') = (\mbox{d}\omega) \, \omega'
    + (-1)^r \omega \, \mbox{d} \omega'
\end{eqnarray}
where $\omega \in \Omega^r({\cal A})$.
\vskip.2cm

The Grassmann algebra of differential forms on
a $C^\infty$-manifold $\cal M$ together with the exterior derivative
operator $\mbox{d}$ constitutes a differential algebra. Here
${\cal A} = C^\infty ({\cal M})$, the (commutative) algebra of
infinitely often differentiable functions on $\cal{M}$. Whereas in this
case differentials and functions commute, this need not be so in general
for a differential calculus on a commutative algebra. Corresponding
examples of {\em non}-commutative differential calculi on commutative
algebras will be discussed in the following sections. In the case of
the algebra of functions on a finite set we are actually forced to
dispense with commutativity of functions and differentials (see
section 5.1).

Given an algebra $\cal A$, there are many different choices of
differential algebras. The problem arises which of them should we choose
to work with. What is the significance of different choices? In
particular, section 5 will provide us with an interesting answer to
this question.

\subsection{The universal differential algebra}
For any associative algebra $\cal A$ one can construct the so-called
{\em universal differential algebra}\footnote{Respectively, {\em
universal differential envelope} of $\cal A$, see \cite{Coqu89} and
references given there.}
in the following way.
We formally associate with every element $a \in {\cal A}$ a symbol
$\mbox{d}a$ and define $\Omega({\cal A})$ to be the
linear span over $\Cx$ of all `words' which can be formed from the
symbols $a$ and $\mbox{d}a$ ($\forall a \in {\cal A}$). Assuming the
Leibniz rule $(\mbox{d}a) \, b = \mbox{d} (a b) - a \, \mbox{d}b \;
(\forall a,b \in {\cal A})$, each element of $\Omega({\cal A})$ can be
written as a sum of monomials of the form $a_0 (\mbox{d}a_1) \cdots
(\mbox{d}a_n)$ with $a_i \in {\cal A}, i=1,\ldots,n$ and $a_0 \in \Cx$
or $a_0 \in {\cal A}$. It is then sufficient to define the action of
$\mbox{d}$ on such monomials:
\begin{eqnarray}
     \mbox{d}(a_0 \, \mbox{d} a_1 \cdots \mbox{d} a_n) =
         \mbox{d} a_0 \, \mbox{d} a_1 \cdots \mbox{d} a_n
\end{eqnarray}
where $\mbox{d}a_0=0$ if $a_0 \in \Cx$.
As a consequence, we have $\mbox{d}^2=0$ and the graded Leibniz rule
(\ref{Leibniz}). The differential algebra obtained in this way
is called {\em universal differential algebra} of $\cal A$.
If $\cal A$ has a  unit element $\idty$ such that $\mbox{d} \idty =0$,
there is a representation in terms of tensor products of
$\cal A$ where $\mbox{d} a = \idty \otimes a - a \otimes \idty$ and,
more generally,\footnote{The argument of $\mbox{d}$ should actually
be restricted to an $n$-form. The tensor product is over $\Cx$.
The space of 1-forms is $\mbox{ker} \, m$ where $m$ is
the multiplication map ${\cal A} \otimes {\cal A} \rightarrow {\cal
A}$.}
\begin{eqnarray}
  \mbox{d}(a_0 \otimes a_1 \otimes \cdots \otimes a_n) =
  \sum_{k=0}^{n+1} (-1)^k \, a_0 \otimes \cdots a_{k-1} \otimes \idty
  \otimes a_k \otimes \cdots \otimes a_n   \; .
\end{eqnarray}
Multiplication is defined by concatenation. For example,
\begin{eqnarray}
 a_0 \, \mbox{d}a_1 \, \mbox{d}a_2 = (a_0 \otimes a_1 - a_0 a_1
 \otimes \idty) \, \mbox{d}a_2
   &=& a_0 \otimes a_1 \otimes a_2 - a_0 \otimes a_1 a_2 \otimes \idty
              \nonumber \\
   & & - a_0 a_1 \otimes \idty \otimes a_2 + a_0 a_1 \otimes a_2
       \otimes \idty     \; .
\end{eqnarray}
Any differential algebra over $\cal A$ can be obtained from the
universal differential algebra as a quotient with respect to some
differential ideal. The universal differential algebra should
therefore be the starting point for a systematic exploration of
differential calculi on a given algebra  $\cal A$ (see section 5).

\section{Differential calculus and lattices}
\setcounter{equation}{0}
The simplest deformation of the ordinary differential calculus on
$\Rl^n$ leads to lattice theories. This is the topic of this section
which is based on \cite{DMH92,DMHS93}.

\subsection{The one-dimensional case}
A deformation of the ordinary differential calculus on
$\Rl$ is obtained as follows. Let $x$ be the identity function on
$\Rl$ (i.e., $x(r) = r \; \forall r \in \Rl$) and $\mbox{d}x$
its (formal) differential. We impose the commutation relation
\begin{eqnarray}            \label{x-dx}
   \lbrack x , \mbox{d} x \rbrack = \ell \, \mbox{d} x
\end{eqnarray}
where $\ell$ is a positive real constant. As a consequence,
\begin{eqnarray}
   f(x) \, \mbox{d} x = \mbox{d} x \, f(x + \ell)
\end{eqnarray}
for a function of $x$. Commuting $f(x)$ with $\mbox{d} x$ thus
results in a discrete translation of the argument of $f$.
For $\cal A$ we take the algebra of all functions $\Rl \rightarrow
\Rl$. They can be regarded as functions of $x$. Let us assume that
$\mbox{d} x$ generates the space $\Omega^1({\cal A})$ of 1-forms
as a right-$\cal A$-module. Then
\begin{eqnarray}
     \mbox{d} f = \mbox{d} x \, \partial_x f
\end{eqnarray}
defines a linear operator $\partial_x \, : \, {\cal A} \rightarrow
{\cal A}$. Now
\begin{eqnarray}
            \ell \, \mbox{d} x \, \partial_x f
  &=& \lbrack x , \mbox{d} x \rbrack \; \partial_x f
   =  \lbrack x , \mbox{d} f \rbrack                    \nonumber \\
  &=& \mbox{d} (x \, f - f \, x ) +  \lbrack f , \mbox{d} x \rbrack
      \qquad \mbox{(using the Leibniz rule for d)}      \nonumber \\
  &=& f(x) \, \mbox{d} x - \mbox{d} x \, f(x) = \mbox{d}x \,
      (f(x+\ell)-f(x))
\end{eqnarray}
shows that
\begin{eqnarray}
  \partial_x f = {1 \over \ell} \, \lbrace f(x+\ell)-f(x) \rbrace \;.
\end{eqnarray}
$\mbox{d} x$ is also a basis of $\Omega^1({\cal A})$ as a
{\em left}-$\cal A$-module, so that
\begin{eqnarray}
 \mbox{d} f =:  \partial_{-x} f \, \mbox{d} x
 = {1 \over \ell} \, \lbrace f(x)-f(x-\ell) \rbrace \, \mbox{d} x \; .
\end{eqnarray}
Hence $\mbox{d}$ acts as a discrete derivative.\footnote{In the limit
$\ell \to 0$, the {\em left-} and {\em right-partial derivatives}
$\partial_{-x}$ and $\partial_x$ tend to the ordinary partial derivative
$\partial/\partial x$ (on differentiable functions).}
In particular, it follows that a `constant' $h$ in the sense that
$\mbox{d} h = 0$ is a function of $x$ with period $\ell$, i.e.
$h(x+\ell) = h(x)$. An indefinite integral associated with $\mbox{d}$
is determined by demanding
\begin{eqnarray}
      \int \mbox{d} f =  f + \mbox{`constant'}   \; .
\end{eqnarray}
For example,
\begin{eqnarray}
 \int \mbox{d} x \; x = \int \mbox{d} x^2 - \int x \, \mbox{d} x
                      = x^2 - \ell \, x - \int \mbox{d} x \; x
\end{eqnarray}
(modulo a `constant') so that
\begin{eqnarray}
  \int \mbox{d} x \; x = {1 \over 2} \, x \, (x-\ell)
                         + \mbox{`constant'}   \; .
\end{eqnarray}
It turns out that every function can be integrated and there is an
explicit formula for its indefinite integral \cite{DMH92}. Since
`constants' with respect to $\mbox{d}$ are $\ell$-periodic functions,
a {\em definite} integral is only defined over intervals of length a
multiple of $\ell$. In this case one finds \cite{DMH92}
\begin{eqnarray}
  \int_{x_0 - m \ell}^{x_0 + n \ell} \mbox{d} x \; f(x)
         = \ell \, \sum_{k=-m}^{n-1} f(x_0 + k \, \ell)
\end{eqnarray}
($m,n \in \Nl$). The integral
\begin{eqnarray}
   \int_{x_0 - \infty}^{x_0 + \infty} \mbox{d} x \; f(x)
     = \sum_{k=-\infty}^{\infty} f(x_0+k \, \ell)
\end{eqnarray}
thus picks out the values of $f$ on a {\em lattice} with $\ell$
as the lattice spacing.

\subsection{Generalization to higher dimensions}
An obvious generalization of (\ref{x-dx}) to $n$ dimensions is
\begin{eqnarray}               \label{x-dx-n}
   \lbrack x^\mu , \mbox{d} x^\nu \rbrack = \ell \, \delta^{\mu \nu}
   \, \mbox{d} x^\nu  \qquad \mbox{(no summation)}  \; .
\end{eqnarray}
As a consequence,
\begin{eqnarray}
   f(x) \, \mbox{d} x^\mu = \mbox{d} x^\mu \, f(x+ \ell^\mu)
\end{eqnarray}
with the notation $(x+\ell^\mu)^\nu := x^\nu + \delta^{\mu \nu} \,
\ell$. Introducing left- and right-partial derivatives via
\begin{eqnarray}
 \sum_{\mu=1}^{n} \partial_{-\mu} f \; \mbox{d} x^\mu
            = \mbox{d} f
            = \sum_{\mu=1}^{n} \mbox{d} x^\mu \; \partial_\mu f
\end{eqnarray}
we find
\begin{eqnarray}
 \partial_{-\mu} f = {1 \over \ell} \, \lbrace f(x)
                     - f(x-\ell^\mu) \rbrace        \qquad \qquad
 \partial_\mu f = {1 \over \ell} \, \lbrace f(x + \ell^\mu)
                  - f(x) \rbrace     \; .
\end{eqnarray}
Hence $\partial_{-\mu}$ and $\partial_\mu$ are discrete (partial)
derivatives.
\vskip.2cm

Again, an indefinite integral is determined by
\begin{eqnarray}
     \int \mbox{d} f = f + \mbox{periodic function}
\end{eqnarray}
and a definite integral is only defined over (a union of) cubes
with edge length $\ell$. In particular, one obtains
\begin{eqnarray}
 \int_{x_0^1}^{x_0^1+\ell} \mbox{d} x^1 \cdots
 \int_{x_0^n}^{x_0^n+\ell} \mbox{d} x^n \, f(x^1, \ldots, x^n)
                = \ell^n \, f(x_0^1, \ldots , x_0^n)   \; .
\end{eqnarray}
\vskip.2cm

Using (\ref{Leibniz}), a consequence of (\ref{x-dx-n}) is
$\mbox{d}x^\mu \, \mbox{d}x^\nu = - \mbox{d}x^\nu \, \mbox{d}x^\mu$.
The familiar anticommutation rule for 1-forms thus holds for the
differentials $\mbox{d}x^\mu$. It does not hold for arbitrary
1-forms, however. Nevertheless, this allows us to introduce a
Hodge operator as follows:
\begin{eqnarray}
 \star \, (\mbox{d}x^{\mu_1} \cdots \mbox{d}x^{\mu_k}) &:=&
 {1 \over (n-k)!} \, \sum_{\mu_{k+1},\ldots,\mu_n=1}^n
 \epsilon^{\mu_1 \ldots \mu_k}{}{}_{\mu_{k+1} \ldots \mu_n} \,
 \mbox{d}x^{\mu_{k+1}} \cdots \mbox{d}x^{\mu_n}    \\
 \star \, ( f(x) \, \omega) &:=& (\star \, \omega) \, f(x)  \; .
\end{eqnarray}
Here we have introduced $\epsilon^{\mu_1 \ldots \mu_n}$ which is
totally antisymmetric with $\epsilon^{1 \ldots n} = 1$. Indices
have been lowered with a Euclidean (or Minkowski) metric with
coefficients $\delta_{\mu \nu}$ (or $\eta_{\mu \nu}$) with respect
to the coordinate functions $x^\mu$.
\vskip.2cm

We now have all the ingrediences for the construction of Lagrangians
and actions in terms of differential forms. For a real scalar field
$\phi$ in $n=4$ dimensions the continuum action can be written as
\begin{eqnarray}
  \int - {1 \over 2} \, (\star \, \mbox{d} \phi) \, \mbox{d} \phi
       + {1 \over 2} \, m^2 \, (\star \, \phi) \, \phi   \; .
\end{eqnarray}
This expression also makes sense for our deformed differential
calculus if the integration is understood over a union of cubes
with edge length $\ell$. It is then easily shown to reproduce the
usual lattice action.
\vskip.2cm

The transition from the continuum to a lattice regularizes the
corresponding quantum field theory. We see that this is achieved
by introducing a noncommutativity between functions and
differentials. We do not need `noncommuting coordinates' (i.e. a
noncommutative algebra $\cal A$) to achieve this. Since divergencies
appear in {\em integrated} expressions, noncommutativity between
differentials and functions seems to be sufficient.

\subsection{Lattice gauge theory}
Let $\bf G$ be a unitary Lie group. A {\em connection} (or `gauge
potential') is a 1-form
\begin{eqnarray}
      A = \sum_{\mu = 1}^n   \mbox{d} x^\mu \; A_\mu
\end{eqnarray}
which transforms according to the familiar rule
\begin{eqnarray}            \label{A-transf}
      A \mapsto G \, A \, G^{-1} - \mbox{d}G \, G^{-1}
\end{eqnarray}
under a gauge transformation with a $\bf G$-valued function $G$.
For $\ell = 0$ the gauge potential is Lie algebra valued. For
$\ell > 0$, however, it is rather group algebra valued (note that
$\mbox{d}$ is a discrete derivative). Then, as a consequence of
(\ref{A-transf}), one finds that
\begin{eqnarray}
      U := \sum_\mu \mbox{d}x^\mu \, U_\mu
        := \sum_\mu \mbox{d}x^\mu - \ell \, A
\end{eqnarray}
transforms homogeneously,
\begin{eqnarray}
  U \mapsto G \, U \, G^{-1} \qquad \qquad
  U_\mu(x) \mapsto G(x+\ell^\mu) \, U_\mu(x) \, G(x)^{-1}   \; .
\end{eqnarray}
The (gauge covariant) field strength of $A$ is
\begin{eqnarray}
  F = \mbox{d}A + A^2 = {1 \over \ell^2} \, \sum_{\mu,\nu}
      \mbox{d}x^\mu \, \mbox{d}x^\nu \, U_\mu(x+\ell^\nu) \, U_\nu(x)
\end{eqnarray}
and the Yang-Mills action in $n$ dimensions becomes
\begin{eqnarray}
  S_{YM} &=& \int \mbox{tr} \lbrack (\star \, F^\dagger) \, F \rbrack
                                \nonumber \\
         &=& \int \mbox{d}^n x \, \mbox{tr} \sum_{\mu,\nu} \ell^{-2n}
             \, \left \lbrack 1 - U_\nu(x)^{-1} U_\mu(x+\ell^\nu)^{-1}
             U_\nu(x+\ell^\mu) U_\mu(x) \right \rbrack
\end{eqnarray}
where we have made the additional assumption that $U_\mu$ has an
inverse in $\bf G$ (which restricts the allowed connections $A$).
When the integral is evaluated over a union of cubes with edge length
$\ell$, the last expression reproduces the Wilson action of lattice
gauge theory.

\subsection{Transformation to $q$-calculus}
Let us return to the one-dimensional (deformed) differential
calculus of section 3.1 and make a transition to a new (coordinate)
function
\begin{eqnarray}
           y = q^{x/\ell}
\end{eqnarray}
with $q \in \Cx \setminus \lbrace 0,1 \rbrace$. The basic commutation
relation (\ref{x-dx}) is then turned into
\begin{eqnarray}
              y \, \mbox{d} y = q \; \mbox{d}y \, y  \; .
\end{eqnarray}
For a function of $y$ we introduce left- and right-partial derivatives
with respect to $y$ via
\begin{eqnarray}
 \partial_{-y} f \; \mbox{d}y = \mbox{d} f = \mbox{d}y \, \partial_y f
                                                        \; .
\end{eqnarray}
A simple calculation reveals that these are the $q$-{\em derivatives}
\begin{eqnarray}
  \partial_{-y} f = { f(y) - f(q^{-1} y) \over (1-q^{-1}) \, y }
     \qquad \qquad
  \partial_y f = { f(q y) - f(y) \over (q-1) \, y }    \; .
\end{eqnarray}
They span a `quantum plane'
\begin{eqnarray}
  \partial_{-y} \, \partial_y = q \, \partial_y \, \partial_{-y}
\end{eqnarray}
and satisfy $q$-deformed canonical commutation relations with $y$
(regarded as a multiplication operator),
\begin{eqnarray}
  \partial_{-y} \, y - q^{-1} \, y \, \partial_{-y} = 1 \qquad \qquad
  \partial_y \, y - q \, y \, \partial_y = 1      \; .
\end{eqnarray}
Furthermore, the integral introduced in section 3.1 is transformed
into the $q$-{\em integral}:
\begin{eqnarray}
 \int_{0-\infty}^{0+\infty} \mbox{d}x \ldots \quad \mapsto \quad
 \int_0^\infty \mbox{d}y \, f(y) = (q-1) \, \sum_{k=-\infty}^{\infty}
  f(q^k) \, q^k    \; .
\end{eqnarray}
Here we have to assume that $q$ is not a root of
unity.\footnote{The case when $q^N = 1$ for some $N \in \Nl$
corresponds to taking $x$ modulo $N \ell$. We are then dealing with
a periodic lattice.}
The $q$-deformed canonical commutation relations which we obtained above
have been considered by various authors. They can be used to formulate
a version of $q$-deformed quantum mechanics (see \cite{DMH92} and the
references therein).

\section{A deformation of the ordinary calculus of differential forms
         on a manifold}
\setcounter{equation}{0}
According to our present knowledge gravity couples to all kind of
matter and cannot be shielded away. This universality suggests to
build it into the most basic structure needed to formulate physical
theories, namely the differential calculus. Along these lines of
thought one is led \cite{DMH92-grav} to a differential algebra with
\begin{eqnarray}                               \label{x-dx-tau}
   \lbrack x^i , \mbox{d}x^j \rbrack = \tau \; g^{ij}
\end{eqnarray}
where $g^{ij}$ (which in the above context should be the space-time
metric) has to be symmetric as required by the Leibniz rule for
$\mbox{d}$. $\tau$ is a 1-form for which we require that
\begin{eqnarray}                               \label{com-tau}
  \lbrack x^i , \tau \rbrack = 0 \qquad \tau^2 = 0 \qquad
  \mbox{d}\tau = 0   \; .
\end{eqnarray}
Furthermore, we choose ${\cal A} = C^\infty({\cal M})$ for a manifold
$\cal M$ and demand that $\mbox{d}x^i$ and $\tau$ form a basis of
$\Omega^1({\cal A})$ as a right- (or left-)
$\cal A$-module.\footnote{This structure is chosen as to cover some
examples to which we will turn in subsections. These examples must
not necessarily be put into the physical context outlined in the
beginning of this section.}
As a consequence, the differential of a function $f$ can be written
as
\begin{eqnarray}
   \mbox{d} f = \tau \; \hat{\partial}_0 f + \mbox{d} x^i \;
                \hat{\partial}_i f       \label{gen-deriv}
\end{eqnarray}
which defines generalized (right-) partial derivative operators
$\hat{\partial}_0, \hat{\partial}_i \, : \, {\cal A} \rightarrow
{\cal A}$.
Using the Leibniz rule for $\mbox{d}$, (\ref{gen-deriv}), the first
of the commutation relations (\ref{com-tau}), $ \lbrack f , \mbox{d}
x^i \rbrack = \lbrack x^i , \mbox{d} f \rbrack$ (using again the Leibniz
rule for $\mbox{d}$) and (\ref{x-dx-tau}), we find
\begin{eqnarray}
 \tau \; \hat{\partial}_0 (f h) + \mbox{d} x^i \;
     \hat{\partial}_i (f h)
 = \tau \, \lbrack (\hat{\partial}_0 f) \, h + f \, \hat{\partial}_0 h
   + g^{ij} (\hat{\partial}_i f) (\hat{\partial}_j h) \rbrack
   + \mbox{d} x^i \, \lbrack (\hat{\partial}_i f) \, h
   + f \, \hat{\partial}_i h \rbrack   \; .
\end{eqnarray}
Hence $\hat{\partial}_i$ is a derivation and therefore a vector field
(since all derivations of $C^\infty ({\cal M})$ are vector fields).
(\ref{gen-deriv}) in particular yields $\mbox{d} x^i = \tau \;
\hat{\partial}_0 x^i + \mbox{d} x^j \; \hat{\partial}_j x^i$ which
implies $\hat{\partial}_0 x^i = 0$ and $\hat{\partial}_j x^i =
\delta^i_j$. Taking this into account, we obtain $\hat{\partial}_i
= \partial_i$ (the ordinary partial derivative). Writing
$\hat{\partial}_0 = {1 \over 2} \, g^{ij} \partial_i \partial_j +
\delta$, we find that $\delta$ must be a derivation and therefore a
vector field. But then $\hat{\partial}_0 x^i = 0$ enforces
$\delta = 0$. Hence
\begin{eqnarray}
   \mbox{d} f = \tau \; {1 \over 2} \, g^{ij} \partial_i \partial_j  f
   + \mbox{d} x^i \; \partial_i f  \; .            \label{df-tau}
\end{eqnarray}
The differential of a function thus involves a {\em second order}
differential operator. Another suprise is that our basic
commutation relation (\ref{x-dx-tau}) is invariant under changes
of coordinates if $g^{ij}$ are tensor components and $\tau$ is
a 1-form:
\begin{eqnarray}
 \lbrack {x'}^i , \mbox{d} {x'}^j \rbrack &=& \lbrack {x'}^i ,
  \mbox{d} x^\ell \, \partial_\ell {x'}^j + \tau \,
   \hat{\partial}_0 {x'}^j \rbrack    \nonumber \\
  &=&  \lbrack {x'}^i , \mbox{d} x^\ell \rbrack \,
       \partial_\ell {x'}^j        \nonumber \\
  &=&  \lbrack x^\ell , \mbox{d} {x'}^i \rbrack \,
       \partial_\ell {x'}^j        \nonumber \\
  &=&  \lbrack x^\ell , \mbox{d} x^k \rbrack \,
       (\partial_k {x'}^i) \, (\partial_\ell {x'}^j)
                                        \nonumber \\
  &=&  \tau \, g^{k \ell} \,
       (\partial_k {x'}^i) \, (\partial_\ell {x'}^j)
                                        \nonumber \\
  &=&  \tau \, {g'}^{k \ell}   \; .
\end{eqnarray}
The deformed differential calculus is therefore well-defined on a
manifold if $g$ and $\tau$ are globally defined on $\cal M$.
This opens a way towards some kind of {\em second order
differential geometry}. Corresponding constructions will be
presented after a formulation of gauge theory with the deformed
differential calculus in the following subsection. What we present
here is not a physical theory yet but rather a mathematical
framework which we expect to be useful for the formulation
and investigation of physical theories.

\subsection{Gauge theory}
Let $\psi$ be an element of ${\cal A}^n$ which transforms as
$\psi \mapsto \psi' = G \, \psi$ under a representation of a Lie
group. For local transformations we can construct a covariant
derivative in the usual way,
\begin{eqnarray}
     D \psi = \mbox{d} \psi + A \, \psi     \; .
\end{eqnarray}
This is indeed covariant if the 1-form $A$ transforms according
to the familiar rule
\begin{eqnarray}             \label{A-trans}
     A' = G \, A \, G^{-1} - \mbox{d} G \, G^{-1}  \; .
\end{eqnarray}
$A$ can then be written in a unique way as
\begin{eqnarray}
     A = \tau \, {1 \over 2} \, A_\tau + \mbox{d} x^i \, A_i  \; .
\end{eqnarray}
Inserting this expression in (\ref{A-trans}), we find that $A_i$
behaves as an ordinary gauge potential and
\begin{eqnarray}
     A_\tau = g^{ij} \, (\partial_i A_j - A_i A_j) + M
\end{eqnarray}
where $M$ is an arbitrary tensorial part ($M' = G M G^{-1}$).
In order to be able to read off gauge-covariant components from
gauge-covariant differential forms, we need the covariantized
differentials $Dx^i := \mbox{d} x^i - \tau \, A^i$.
They transform as $D'x^i = G \, Dx^i \, G^{-1}$.
The covariant derivative of $\psi$ can now be written as
\begin{eqnarray}
     D \psi = \tau \, {1 \over 2} \, (g^{ij} D_i D_j + M) \, \psi
              + Dx^i \, D_i \psi
\end{eqnarray}
where $D_i$ denotes the ordinary covariant derivative (using $A_i$).
The field strength of $A$ is
\begin{eqnarray}
   F  = \mbox{d} A + A^2
      = \tau \, {1 \over 2} \, (\mbox{d} x^i \, D^j F_{ji} - DM)
         + {1 \over 2} \, Dx^i \, Dx^j \, F_{ij}
\end{eqnarray}
which involves the Yang-Mills operator (when $g^{ij}$ is identified
with the space-time metric).
$F_{ij}$ is the (ordinary) field strength of $A_i$.

\subsection{Generalized differential geometry}
For a vector field $Y^i$ we introduce a (right-) covariant derivative
\begin{eqnarray}
           D Y^i := \mbox{d} Y^i + Y^j \; {}_j \Gamma^i   \; .
\end{eqnarray}
This is indeed {\em right}-covariant iff the generalized connection
${}_j \Gamma^i$ is given by
\begin{eqnarray}
   {}_j \Gamma^i = \tau \, {1 \over 2} \, \left \lbrack
                   g^{k \ell} ( \partial_k \Gamma^i{}_{j \ell}
                   + \Gamma^i{}_{m k} \Gamma^m{}_{j \ell} ) + M^i{}_j
                                          \right \rbrack
                   + \mbox{d} x^k \, \Gamma^i{}_{jk}
\end{eqnarray}
where $\Gamma^i{}_{jk}$ are the components of an ordinary linear
connection on $\cal M$ and $M^i{}_j$ is a tensor.
\vskip.2cm

The coordinate differentials $\mbox{d} x^i$ do not transform
covariantly, since
\begin{eqnarray}
 \mbox{d}{x'}^k = \tau \, {1 \over 2} \, g^{ij} \partial_i \partial_j
             {x'}^k + \mbox{d} x^\ell \, \partial_\ell {x'}^k
\end{eqnarray}
as a consequence of (\ref{df-tau}). But the 1-forms
\begin{eqnarray}
 D x^k := \mbox{d} x^k + \tau \, {1 \over 2} \, \Gamma^k{}_{ij}
          \, g^{ij}
\end{eqnarray}
are {\em right}-covariant and (\ref{df-tau}) can now be rewritten as
\begin{eqnarray}
    \mbox{d} f = \tau \, {1 \over 2} \, g^{ij} \nabla_i \nabla_j f
                 + D x^i \; \partial_i f
\end{eqnarray}
where $\nabla_i$ denotes the ordinary covariant derivative (with
$\Gamma^i{}{}_{k \ell}$).
Also the covariant exterior derivative of $Y^i$ can now be written in
an explicitly right-covariant form,
\begin{eqnarray}
  D Y^i = \tau \, {1 \over 2} \, (g^{k \ell} \nabla_k
               \nabla_\ell Y^i + M^i{}_j \, Y^j)
               + D x^j \; \nabla_j Y^i   \; .
\end{eqnarray}
The product of differentials $D x^k$ is neither antisymmetric nor
covariant, but
\begin{eqnarray}
  D x^k \wedge D x^\ell := D x^k \, D x^\ell -
  \tau \, \mbox{d} x^j \, g^{i \ell} \, \Gamma^k_{ij}
\end{eqnarray}
is antisymmetric and right-covariant. Then
\begin{eqnarray}            \label{gen-torsion}
  D^2 x^k = \Theta^k
            - \tau \, {1 \over 2} \, \mbox{d} x^\ell \, ( R^k{}_\ell -
            M^k{}_\ell - \nabla^j Q^k{}_{j \ell} + Q^{kij} \,
            Q_{ij \ell} )
\end{eqnarray}
with the torsion 2-form
\begin{eqnarray}
  \Theta^k := {1 \over 2} \, D x^i \wedge D x^j \, Q^k{}_{ij}
\end{eqnarray}
where $Q^k{}_{ij}$ is the torsion of $\Gamma^k{}_{ij}$.
(\ref{gen-torsion}) may be regarded as a generalized torsion. Its
vanishing implies the vanishing of $Q^k{}_{ij}$ and an
equation of the form of Einstein's equations. It suggests to relate
$M^i{}_j$ to the energy-momentum tensor of matter. The latter is
then `geometrized' in this formalism as part of the generalized
connection.
\vskip.2cm

We have seen in particular that the (covariant) exterior derivative
of a field contains in its $\tau$-part the corresponding part of the
field equation to which it is usually subjected in physical models.
We refer to \cite{DMH92-grav} for an ample discussion.

\subsection{Example: Classical limit of bicovariant differential
            calculus on a quantum group}
The Hopf algebra structure of a quantum group can be used to narrow
down the large number of possible differential calculi on such an
algebra and leads to the concept of `bicovariant differential calculus'
\cite{Woro89}. One is mainly interested in such calculi where
the dimension of the space of 1-forms (as a left or right $\cal
A$-module) coincides with the number of generators of the
quantum group.

The quantum group $GL_q(2)$ is an algebra $\cal A$ generated by
elements $x^\mu, \, \mu=1,\ldots,4$, which satisfy certain commutation
relations. All possible bicovariant differential calculi on
$GL_q(2)$ were found in \cite{MH92} (see also \cite{MH+Reut93}).
They form a 1-parameter family (with a complex parameter $s$).
In the classical limit $q \to 1$ one obtains \cite{DMH93-stoch}
\footnote{The limit was obtained by regarding the parameter $s =
A^1_4$ in the notation of \cite{MH92,MH+Reut93} as independent of
$q$. In this parametrization there are two calculi for a given value of
$s$. Here we refer to the one for which $A^1_1 = 1+s$ (in the
notation of the cited papers). After a $q$-dependent transformation
to another parameter $t$ (as in \cite{MH93-Claus}) we may as well regard
this new parameter $t$ as $q$-independent. Then the limit $q \to 1$
yields in general a different result. This is in fact the case with
the parameter $t$ chosen in \cite{MH93-Claus}. The statement made
there about how the limit was taken is incorrect.}
\begin{eqnarray}
 \lbrack x^\mu , \mbox{d} x^\nu \rbrack = \tau \; \gamma^{\mu \nu}
\end{eqnarray}
with
\begin{eqnarray}
 \gamma^{\mu \nu} = -(x^1 x^4-x^2 x^3)^{-1} \, x^\mu x^\nu
    + 4 \, (\delta^{(\mu}_1 \delta^{\nu)}_4 - \delta^{(\mu}_2
    \delta^{\nu)}_3)    \\
  \tau = s \, (x^4 \, \mbox{d} x^1 - x^3 \, \mbox{d} x^2
         - x^2 \, \mbox{d} x^3 + x^1 \, \mbox{d} x^4)
       =: \tau_\nu \, \mbox{d} x^\nu   \; .
\end{eqnarray}
These objects satisfy $\gamma^{\mu \nu} \, \tau_\nu = 0$ which
defines a Galilei structure\footnote{This is a
generalization of the Newtonian space-time in which case $\gamma^{\mu
\nu}$ is the space-metric and $\tau = \mbox{d}t$ where $t$ is the
absolute time.}.

Passing from differential calculus on $GL_q(2)$ to differential
calculus on $SL_q(2,\Rl)$ one has to fix the parameter $s$. From the
two resulting bicovariant differential calculi only one has a
reasonable classical limit. After elimination of $x^4$ with the
help of the determinant constraint in a coordinate patch where
$x^1 \neq 0$, we obtain \cite{MH+Reut93}
\begin{eqnarray}
   \lbrack x^i , \mbox{d}x^j \rbrack = \tau \; g^{ij}
   \qquad (i,j = 1,2,3)
\end{eqnarray}
where $g^{ij}$ is the bi-invariant metric on $SL(2,\Rl)$. The 1-form
$\tau$ cannot be expressed as $\sum_i f_i(x^j) \, \mbox{d} x^i$ or
$\sum_i \mbox{d} x^i \, f_i(x^j)$. The 1-forms $ \mbox{d} x^i$ and
$\tau$ thus form a basis of $\Omega^1({\cal A})$.
We refer to \cite{MH+Reut93,DMH93-stoch,MH93-Claus} for further details
and to \cite{DMH93-Mex} for a generalization of the above example.

\subsection{Relation with `proper time' theories
            and stochastic calculus on manifolds}
When $\tau = \gamma \, \mbox{d} t \, $ with a constant $\gamma$, we may
consider (smooth) functions $f(x^i,t)$ depending also on the parameter
$t$. (\ref{df-tau}) then has to be replaced by
\begin{eqnarray}        \label{df-dt}
  \mbox{d}f = \mbox{d}t \, (\partial_t + {\gamma \over 2} \, g^{ij} \,
  \partial_i \partial_j ) \, f + \mbox{d}x^i \, \partial_i f   \; .
\end{eqnarray}
1. Let $\gamma = - i \, \hbar$. The requirement that the
$\mbox{d} t$-part vanishes leads to the Schr\"odinger equation
\begin{eqnarray}
  (i \, \hbar \, \partial_t + {\hbar^2 \over 2} \, g^{ij} \,
  \partial_i \partial_j ) \, f = 0   \; .
\end{eqnarray}
If $g^{ij}$ is a space-time metric, this is the five-dimensional
Schr\"odinger equation of `proper time' quantum theory ({\`a} la
Fock, Stueckelberg, Nambu, Feynman, \ldots, see
\cite{Fanc93} for a recent review). The noncommutative differential
calculus may be viewed as a basic structure underlying such
proper time theories (see \cite{DMH92-grav} for an extensive
discussion).
\vskip.2cm
\noindent
2. The formula (\ref{df-dt}) -- with a positive definite metric $\gamma
\, g^{ij}$ -- is well-known in the theory of stochastic processes
(It\^o calculus) and suggests that the noncommutative
differential calculus provides us with a convenient framework to
deal with stochastic processes on manifolds (cf \cite{Nels85} where
$\gamma = \hbar$ and $g$ is the inverse mass matrix). There is
indeed a translation \cite{DMH93-stoch} to the (It\^o) calculus of
(commutative) stochastic differentials where one has the following
rules for products of stochastic differentials:
\begin{eqnarray}            \label{Ito}
 \mbox{d}t \bullet \mbox{d}t = 0, \qquad
 \mbox{d}t \bullet \mbox{d}x^i_t = 0, \qquad
 \mbox{d}x^i_t \bullet \mbox{d}x^j_t = \gamma \, g^{ij} \, \mbox{d}t
                                            \; .
\end{eqnarray}
The coordinates $x^i$ have been replaced by stochastic processes
$x^i_t$ and $g^{ij}$ is a function of the latter. In the stochastic
calculus one has a deformed Leibniz rule:
\begin{eqnarray}
  \mbox{d} (f_t \, h_t) = (\mbox{d} f_t) \, h_t + f_t \, \mbox{d}
  h_t + (\mbox{d} f_t) \bullet \mbox{d} h_t    \; .
\end{eqnarray}
But it is possible to define a new (noncommutative) product $\ast$
between functions and differentials such that
\begin{eqnarray}
 \mbox{d} (f_t \, h_t) = (\mbox{d} f_t) \ast h_t + f_t \ast \mbox{d}
  h_t     \; .
\end{eqnarray}
(This product is related to the Moyal product.) In terms of the
corresponding commutator the relations (\ref{Ito}) are translated into
\begin{eqnarray}
 \lbrack t, \mbox{d} t \rbrack_\ast = 0 \, , \quad
 \lbrack t, \mbox{d} x^i_t \rbrack_\ast = \lbrack x^i_t, \mbox{d} t
                                          \rbrack_\ast = 0 \, , \quad
 \lbrack x^i_t, \mbox{d} x^j_t \rbrack_\ast = \gamma \, g^{ij}
                                              \, \mbox{d}t  \; .
\end{eqnarray}
In contrast to the It\^o calculus, the noncommutative
differential calculus admits an extension to forms of higher grade.
Do they have a stochastic interpretation?
We refer to \cite{DMH93-stoch,DMH93-Mex} for further details. The
formalism should also be of interest in the context of stochastic
quantization.

\section{Some aspects of differential calculus on discrete sets}
\setcounter{equation}{0}
When one formulates differential calculus on (the commutative algebra
of functions on) a finite or, more generally, discrete set,
functions and differentials turn out to be necessarily
{\em non}commutative. The simplest example is provided by Connes'
2-point model which received a lot of interest as an ingredient of
particle physics models \cite{Conn+Lott90}. Another example, a lattice
differential calculus, was already discussed in section 3. These two
examples motivated a general study of differential calculi on arbitrary
discrete sets. This has been carried out in some recent papers
\cite{DMH93-fs,DMH94} (see also \cite{Sita92} for the case of discrete
groups). We believe that this kind of mathematics is very convenient
for discrete modelling and in particular for developping ideas of
discrete space-time (cf \cite{Sork91}).

\subsection{Differential calculus on discrete sets}
Let $\cal M$ be a discrete\footnote{In the case of {\em infinite} sets
the following calculus is `formal' and more work has to be done to put
it on a rigorous footing.}
set of elements $i,j, \ldots$ and $\cal A$
the algebra of $\Cx$-valued functions on it. With each element $i \in
{\cal M}$ we associate a function $e_i \in {\cal A}$ via
\begin{eqnarray}
      e_i(j) = \delta_{ij}      \; .
\end{eqnarray}
Then $e_i \, e_j = \delta_{ij} \, e_j$ and $\sum_i e_i = \idty$ where
$\idty(i) = 1 \; \forall i \in {\cal M}$. Acting with $\mbox{d}$ on
these relations and using the Leibniz rule yields
\begin{eqnarray}
  e_i \, \mbox{d} e_j = - \mbox{d} e_i \; e_j + \delta_{ij} \,
                        \mbox{d} e_j
              \qquad \qquad
  \sum_i \mbox{d} e_i = 0
\end{eqnarray}
where the first formula expresses the {\em non}commutativity of the
differential calculus. It turns out to be convenient to introduce
the following 1-forms,
\begin{eqnarray}
     e_{ij} := e_i \, \mbox{d} e_j    \qquad  (i \neq j)
\end{eqnarray}
(we set $e_{ii} = 0$). From these 1-forms we build the $(r-1)$-forms
\begin{eqnarray}            \label{e...}
  e_{i_1 \ldots i_r} := e_{i_1 i_2} \, e_{i_2 i_3} \cdots
                        e_{i_{r-1} i_r}   \; .
\end{eqnarray}
Now one has the following formulae,
\begin{eqnarray}
  \mbox{d} e_i &=& \sum_k (e_{ki} - e_{ik})  \qquad
  \mbox{d} f    = \sum_{i,j} e_{ij} \, \lbrack f(j) - f(i) \rbrack \\
  \mbox{d} e_{ij} &=& \sum_k (e_{kij} - e_{ikj} + e_{ijk})    \; .
                    \label{de_ij}
\end{eqnarray}
If no further relations are imposed, we are dealing with the
universal differential calculus. In this case the $e_{i_1 \ldots i_r}$
constitute a basis over $\Cx$ of $\Omega^{r-1}({\cal A})$ for $r>1$.

An {\em inner product} on $\Omega({\cal A})$ should have the
properties
\begin{eqnarray}
  ( \psi_r , \phi_s ) &=& 0  \qquad \mbox{if } r \neq s   \\
  ( \psi , c \, \phi + \chi) &=& c \, ( \psi , \phi)
  + (\psi , \chi)      \qquad \forall c \in  \Cx   \\
  \overline{(\psi , \phi)} &=& (\phi , \psi)
\end{eqnarray}
(where $\psi_r \in \Omega^r({\cal A})$ and the bar denotes complex
conjugation). An inner product is then determined by the values
of $(e_{i_1 \ldots i_r}, e_{j_1 \ldots j_r})$. We will require
that this has the structure
\begin{eqnarray}
 (e_{i_1 \ldots i_r}, e_{j_1 \ldots j_r}) = \delta_{i_1 j_1} \,
 g_{i_1 \ldots i_r j_1 \ldots j_r} \, \delta_{i_r j_r}
                             \label{ip-condition}
\end{eqnarray}
with constants $g_{i_1 \ldots i_r j_1 \ldots j_r}$. For the universal
$\Omega({\cal A})$, the latter may be taken proportional to
$\delta_{i_1 j_1} \cdots \delta_{i_r j_r} \, \prod_{s=1}^{r-1}
(1-\delta_{i_s i_{s+1}})$ (the last factor takes care of the fact
that $e_{ii} = 0$). Given an inner product, we can construct
an adjoint $\mbox{d}^\dagger$ of $\mbox{d}$, a Laplace-Beltrami
operator etc..

An {\em involution} on $\Omega({\cal A})$ should have the properties
\begin{eqnarray}
   (\psi \, \phi)^\ast = \phi^\ast \, \psi^\ast   \qquad
   (\mbox{d} \psi_r)^\ast = (-1)^r \, \mbox{d} (\psi_r^\ast) \; .
\end{eqnarray}
If $^\ast$ restricted to $\Omega^0({\cal A}) = {\cal A}$ is complex
conjugation, then
\begin{eqnarray}
     e_{k \ell}^\ast = - e_{\ell k}  \; .    \label{e-involution}
\end{eqnarray}

\subsection{Reductions of the universal differential algebra, graphs
            and topologies}
Smaller differential algebras are obtained from the universal one
by setting some of the $e_{ij}$ to zero (`{\em reduction}'). Let us
associate a {\em graph} with a differential calculus on $\cal M$
in the following way. The vertices correspond to the elements of
$\cal M$ and there is an arrow from $i$ to $j$ iff $e_{ij} \neq 0$.
\vskip.2cm \noindent
{\em Example 1.} The universal differential calculus on a 3-point
set is represented by the graph

\begin{minipage}{9cm}
\unitlength2.cm
\begin{picture}(4.,1.8)(-2.,-0.4)
\thicklines
\put(0.,1.) {\circle*{0.1}}
\put(1.,0.) {\circle*{0.1}}
\put(-1.,0.) {\circle*{0.1}}
\put(-0.8,-0.07) {\vector(1,0){1.6}}
\put(0.8,0.05) {\vector(-1,0){1.6}}
\put(0.8,0.1) {\vector(-1,1){0.7}}
\put(0.2,0.9) {\vector(1,-1){0.7}}
\put(-0.1,0.8) {\vector(-1,-1){0.7}}
\put(-0.9,0.2) {\vector(1,1){0.7}}
\end{picture}
\end{minipage}
\hfill
\begin{minipage}{6cm}
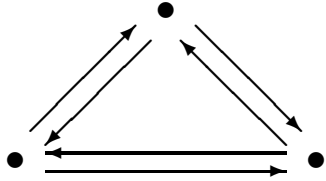
\centerline{\bf Fig.1}
\small
\noindent
The graph associated with the universal differential algebra on a set
with three elements.
\end{minipage}

\vskip.2cm \noindent
{\em Example 2.} Let us consider the differential calculus on the
3-point space associated with the following graph.

\begin{minipage}{9cm}
\unitlength2.cm
\begin{picture}(4.,1.8)(-2.,-0.4)
\thicklines
\put(0.,1.) {\circle*{0.1}}
\put(1.,0.) {\circle*{0.1}}
\put(-1.,0.) {\circle*{0.1}}
\put(-1.25,-0.05) {$0$}
\put(1.2,-0.05) {$1$}
\put(-0.05,1.2) {$2$}
\put(-1.,0.) {\vector(1,0){1.9}}
\put(1.,0.) {\vector(-1,1){0.9}}
\put(0.,1.) {\vector(-1,-1){0.9}}
\end{picture}
\end{minipage}
\hfill
\begin{minipage}{6cm}
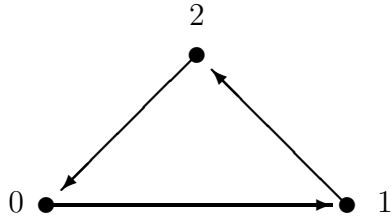
\centerline{\bf Fig.2}
\small
\noindent
The graph associated with a reduction of the universal
differential algebra on a set of three elements.
\end{minipage}
The nonvanishing basic 1-forms are thus $e_{01}, e_{12}, e_{20}$.
The only basic 2-forms we can construct from these are $e_{012}$,
$e_{120}$ and $e_{201}$ according to (\ref{e...}). But
$e_{10}=e_{02}=e_{21}=0$ implies
\begin{eqnarray}
  0 = \mbox{d} e_{10} = \sum_{k=1}^3 (e_{k10} - e_{1k0} + e_{10k})
    = - e_{120}
\end{eqnarray}
and also $e_{012} = e_{201} =0$ so that there are no 2-forms.
Hence we can assign the {\em dimension} 1 to the
3-point set. Let us associate new vertices with the nonvanishing
basic 1-forms and draw an extended graph (Fig.3).

\begin{minipage}[t]{8cm}
\unitlength2.cm
\begin{picture}(4.,1.8)(-2.,-0.4)
\thicklines
\put(-1.,0.) {\circle*{0.1}}
\put(1.,0.) {\circle*{0.1}}
\put(0.,1.) {\circle*{0.1}}
\put(-0.5,0.5) {\circle{0.1}}
\put(0.5,0.5) {\circle{0.1}}
\put(0.,0.) {\circle{0.1}}
\put(-1.25,-0.05) {$0$}
\put(1.2,-0.05) {$1$}
\put(-0.05,1.2) {$2$}
\put(-.85,0.5) {$20$}
\put(0.7,0.5) {$12$}
\put(-0.1,-.3) {$01$}
\put(-1.,0.) {\vector(1,0){0.9}}
\put(0.1,0.) {\vector(1,0){0.8}}
\put(1.,0.) {\vector(-1,1){0.4}}
\put(0.4,0.6) {\vector(-1,1){0.35}}
\put(0.,1.) {\vector(-1,-1){0.4}}
\put(-0.6,0.4) {\vector(-1,-1){0.35}}
\end{picture}
\end{minipage}
\hfill
\begin{minipage}[t]{8cm}
\unitlength2.cm
\begin{picture}(4.,1.8)(-1.,-0.4)
\thicklines
\put(-1.,1.) {\circle*{0.1}}
\put(0.,1.) {\circle*{0.1}}
\put(1.,1.) {\circle*{0.1}}
\put(-1.04,1.2) {$0$}
\put(-0.04,1.2) {$1$}
\put(0.96,1.2) {$2$}
\put(1.4,0.95) {\mbox{0-forms}}
\put(-1.,0.) {\circle{0.1}}
\put(0.,0.) {\circle{0.1}}
\put(1.,0.) {\circle{0.1}}
\put(1.4,-0.05) {\mbox{1-forms}}
\put(-1.1,-0.3) {$01$}
\put(-0.1,-0.3) {$12$}
\put(0.9,-0.3) {$20$}
\put(-1.,1.) {\vector(0,-1){0.9}}
\put(0.,1.) {\vector(0,-1){0.9}}
\put(1.,1.) {\vector(0,-1){0.9}}
\put(-1.,0.) {\vector(1,1){0.9}}
\put(0.,0.) {\vector(1,1){0.9}}
\put(1.,0.) {\vector(-2,1){1.9}}
\end{picture}
\end{minipage}
\vskip.1cm
\centerline{
\begin{minipage}[t]{13cm}
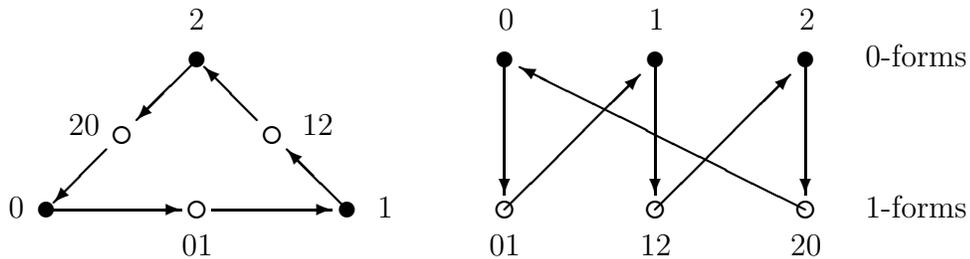
\centerline{\bf Fig.3}
\vskip.2cm \noindent
\small
The extension of the graph in Fig.2 drawn in two different ways. The
second version suggests an interpretation as an (oriented) Hasse
diagram.
\end{minipage}   }

\vskip.3cm
\noindent
Following Sorkin \cite{Sork91}, the second graph in Fig.3 can be
interpreted as a {\em Hasse diagram} which determines a topology
on the 3-point space in the following way.
A vertex together with all lower lying
vertices which are connected to it forms an open set. In the present
case, $\lbrace 01 \rbrace, \lbrace 12 \rbrace, \lbrace 20 \rbrace,
\lbrace 0,01,20 \rbrace, \lbrace 1, 01, 12 \rbrace,
\lbrace 2, 12, 20 \rbrace$  are the open sets (besides the empty
and the whole set). This is an approximation to the topology of $S^1$.
It consists of a chain of three open sets covering $S^1$ which already
displays the global topology of $S^1$. In particular, the fundamental
group $\pi_1$ is the same as for $S^1$.
\vskip.2cm \noindent
{\em Example 3.} On a 4-point set let us consider the differential
calculus associated with the following graph.

\begin{minipage}{9cm}
\unitlength2.cm
\begin{picture}(4.,1.8)(-2.,-0.4)
\thicklines
\put(0.,0.) {\circle*{0.1}}
\put(1.,0.) {\circle*{0.1}}
\put(0.,1.) {\circle*{0.1}}
\put(1.,1.) {\circle*{0.1}}
\put(-0.2,-0.05) {$0$}
\put(1.2,-0.05) {$1$}
\put(1.2,0.95) {$2$}
\put(-.2,0.95) {$3$}
\put(0.,0.) {\vector(1,0){0.9}}
\put(1.,0.) {\vector(0,1){0.9}}
\put(0.,0.) {\vector(0,1){0.9}}
\put(0.,1.) {\vector(1,0){0.9}}
\end{picture}
\end{minipage}
\hfill
\begin{minipage}{6cm}
\centerline{\bf Fig.4}
\small
\noindent
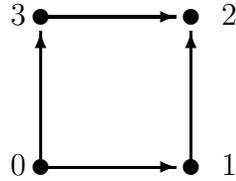
A graph representing a differential calculus on a set of four elements.
\end{minipage}
\vskip.2cm
\noindent
The nonvanishing basic 1-forms are $e_{01}, e_{12}, e_{03}, e_{32}$.
{}From these we can only construct the basic 2-forms $e_{012}$ and
$e_{032}$. No forms of higher grade are present. Furthermore,
(\ref{de_ij}) together with $e_{02}=0$ implies $e_{032} = - e_{012}$
and there remains only one independent basic 2-form (which we will
take to be $e_{012}$). The differential calculus thus assigns
2 dimensions to the 4-point set.
Again, we extend the graph by adding new
vertices corresponding to the nonvanishing basic 1- and 2-forms
(Fig.5).
\vskip.2cm \noindent
\begin{minipage}[t]{8cm}
\unitlength1.5cm
\begin{picture}(4.,2.8)(-2.,-1.4)
\thicklines
\put(-1.,1.) {\circle*{0.1}}
\put(0.,1.) {\circle{0.1}}
\put(1.,1.) {\circle*{0.1}}
\put(-1.04,1.2) {$3$}
\put(-0.08,1.2) {$32$}
\put(0.96,1.2) {$2$}
\put(-1.,0.) {\circle{0.1}}
\put(0.,0.) {\circle{0.1}}
\put(0.,0.) {\circle*{0.04}}
\put(1.,0.) {\circle{0.1}}
\put(-1.4,-0.1) {$03$}
\put(0.1,0.1) {$012$}
\put(1.2,-0.1) {$12$}
\put(-1.,-1.) {\circle*{0.1}}
\put(0.,-1.) {\circle{0.1}}
\put(1.,-1.) {\circle*{0.1}}
\put(-1.04,-1.3) {$0$}
\put(-0.08,-1.3) {$01$}
\put(0.96,-1.3) {$2$}
\put(-1.,-1.) {\vector(1,0){0.9}}
\put(0.1,-1.) {\vector(1,0){0.8}}
\put(-0.9,0.) {\vector(1,0){0.8}}
\put(0.1,0.) {\vector(1,0){0.8}}
\put(-1.,1.) {\vector(1,0){0.9}}
\put(0.1,1.) {\vector(1,0){0.8}}
\put(-1.,-1.) {\vector(0,1){0.9}}
\put(-1.,0.1) {\vector(0,1){0.8}}
\put(1.,-1.) {\vector(0,1){0.9}}
\put(1.,0.1) {\vector(0,1){0.8}}
\put(0.,0.9) {\vector(0,-1){0.8}}
\put(0.,-0.1) {\vector(0,-1){0.8}}
\end{picture}
\end{minipage}
\hfill
\begin{minipage}[t]{8cm}
\unitlength1.5cm
\begin{picture}(4.,2.8)(-0.5,-1.4)
\thicklines
\put(-1.,1.) {\circle*{0.1}}
\put(0.,1.) {\circle*{0.1}}
\put(1.,1.) {\circle*{0.1}}
\put(2.,1.) {\circle*{0.1}}
\put(2.4,1.) {\mbox{0-forms}}
\put(-1.04,1.2) {$0$}
\put(-0.04,1.2) {$1$}
\put(0.96,1.2) {$2$}
\put(1.96,1.2) {$3$}
\put(-1.,0.) {\circle{0.1}}
\put(0.,0.) {\circle{0.1}}
\put(1.,0.) {\circle{0.1}}
\put(2.,0.) {\circle{0.1}}
\put(2.4,0.) {\mbox{1-forms}}
\put(-1.4,-0.1) {$01$}
\put(-0.4,-0.1) {$12$}
\put(0.6,-0.1) {$32$}
\put(1.9,-0.3) {$03$}
\put(0.5,-1.) {\circle{0.1}}
\put(0.5,-1.) {\circle*{0.04}}
\put(0.3,-1.3) {$012$}
\put(2.4,-1.) {\mbox{2-forms}}
\put(-1.,1.) {\vector(0,-1){0.9}}
\put(0.,1.) {\vector(0,-1){0.9}}
\put(-1.,1.) {\vector(3,-1){2.9}}
\put(2.,1.) {\vector(-1,-1){0.9}}
\put(-0.9,0.1) {\vector(1,1){0.8}}
\put(0.1,0.1) {\vector(1,1){0.8}}
\put(1.,0.1) {\vector(0,1){0.8}}
\put(2.,0.1) {\vector(0,1){0.8}}
\put(0.95,-0.1) {\vector(-1,-2){0.4}}
\put(1.85,-0.1) {\vector(-3,-2){1.25}}
\put(0.35,-0.9) {\vector(-3,2){1.3}}
\put(0.45,-0.9) {\vector(-1,2){.4}}
\end{picture}
\end{minipage}

\centerline{
\begin{minipage}[t]{12cm}
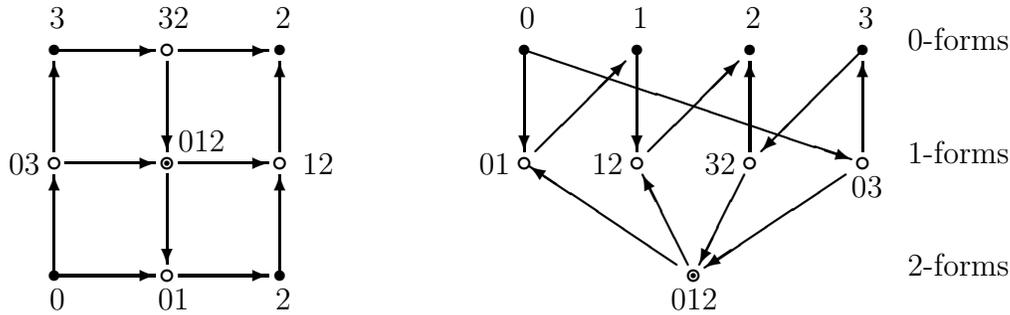
\centerline{\bf Fig.5}
\vskip.1cm \noindent
\small
The extended graph and (oriented) Hasse diagram derived from the
graph in Fig.4.
\end{minipage}    }

\vskip.2cm \noindent
The arrows are determined by the equations
\begin{eqnarray}
  \mbox{d} e_{01} = \mbox{d} e_{12} = - \mbox{d} e_{03}
                  = - \mbox{d} e_{32} = e_{012}  \; .
\end{eqnarray}
For example, $e_{012}$ appears in the expression for
$\mbox{d}e_{01}$. So we connect the corresponding vertices. The
orientation of the arrow is determined by the sign in front of
$e_{012}$. For the topology the orientation is irrelevant, however.
The latter can be visualized as follows (Fig.6).

\begin{minipage}{9cm}
\unitlength2.cm
\begin{picture}(4.,1.8)(-2.,-0.7)
\thicklines
\put(-0.2,0.) {\circle{2.}}
\put(0.2,0.) {\circle{2.}}
\put(-0.2,0.3) {\circle{2.}}
\put(0.2,0.3) {\circle{2.}}
\end{picture}
\end{minipage}
\hfill
\begin{minipage}{6cm}
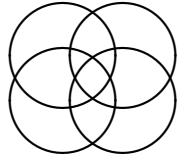
\centerline{\bf Fig.6}
\small
\noindent
The topology on the 4-point set determined by the Hasse
diagram in Fig.5. The four disks and their intersections
represent 2-dimensional open sets.
\end{minipage}

\vskip.2cm \noindent
Further examples can be found in \cite{DMH94}.
\vskip.2cm \noindent
{\em Remark.} In a recent work
\cite{BBET93} the authors start with a topology (e.g., given by
imprecise space-time measurements), construct the Hasse diagram,
assign incidence numbers $\pm 1$ to its edges, construct a
boundary operator etc.. Using the results outlined above, one
recovers a noncommutative differential calculus (on a finite set)
behind all this.                          \hfill  {\Large $\Box$}

\subsection{Gauge fields}
As an example of field theory on a discrete set we consider (pure)
gauge theory. The structures introduced in the following are defined
for any choice of a differential calculus on a discrete set $\cal M$.
A {\em connection 1-form} is an element $A = \sum_{i,j} e_{ij} \,
A_{ij} \, \in \Omega^1({\cal A}) \otimes_{\cal A} M_n({\cal A})$ (where
$M_n({\cal A})$ is the space of $n \times n$ matrices with entries in
$\cal A$) with the familiar transformation rule\footnote{Note that
$A$ cannot be Lie algebra valued since $dG$ is a discrete
derivative. See also section 3.3.}
\begin{eqnarray}
    A \mapsto G \, A \, G^{-1} - \mbox{d} G \; G^{-1}
\end{eqnarray}
where $G = \sum_i G(i) \, e_i$ is an element of a local gauge group,
a subgroup of $GL(n,{\cal A})$. Associated with $A$ is the
{\em transport operator}
\begin{eqnarray}
    U = \sum_{i,j} e_{ij} \, U_{ij} \, ,  \qquad
    U_{ij} := {\bf 1} + A_{ij}
\end{eqnarray}
which transforms as follows,
\begin{eqnarray}
    U \mapsto G \, U \, G^{-1} \qquad
    U_{ij} \mapsto G(i) \, U_{ij} \, G(j)^{-1}  \; .
\end{eqnarray}
The curvature (or field strength) of $A$ is
\begin{eqnarray}
   F := \mbox{d} A + A^2 = \sum_{i,j,k} e_{ijk} \,
        (U_{ij} U_{jk} - U_{ik})   \; .
\end{eqnarray}
It transforms in the familiar way, $F \mapsto G \, F \, G^{-1}$.
In order to generalize an inner product (with the property
(\ref{ip-condition})) to matrix valued forms, we require
that\footnote{Here $\phi_{i_1 \ldots i_r}$ is a matrix with entries
in $\Cx$ and $\phi^\dagger_{i_1 \ldots i_r}$ denotes the hermitian
conjugate matrix.}
\begin{eqnarray}
   (\phi , \psi) = \sum \phi_{i_1 \cdots i_r}^\dagger \,
   (e_{i_1 \cdots i_r}, e_{j_1 \cdots j_s}) \; \psi_{j_1\cdots j_s}
                                           \; .
\end{eqnarray}
The {\em Yang-Mills action}
\begin{eqnarray}
                S_{YM} := \mbox{tr} \, (F,F)
\end{eqnarray}
is then gauge-invariant if $G^\dagger = G^{-1}$. Covariant
derivatives of fields on $\cal M$ can be introduced in the
usual way \cite{DMH94}.

\subsection{How to recover the usual lattice calculus}
Choose ${\cal M} = \Ir^n = \lbrace a = (a^\mu) \, | \;
 \mu = 1, \ldots, n \, , \, a^\mu \in \Ir \rbrace$ and consider the
following reduction of the universal differential calculus on $\cal
 M$ :
\begin{eqnarray}
 e_{ab} \neq 0  \qquad \Leftrightarrow \qquad  b = a + \hat{\mu}
 \; \mbox{ for some } \mu
\end{eqnarray}
where $\hat{\mu} := (\delta^\nu_\mu) \in {\cal M}$.
The corresponding graph is an oriented lattice in $n$ dimensions
(which locally looks like the graph in Fig.4). In terms of the $n$
functions
\begin{eqnarray}
     x^\mu := \ell \, \sum_a a^\mu \, e_a
\end{eqnarray}
(where $\ell$ is a positive constant) one finds the
relation\footnote{Every $f \in {\cal A}$ can be regarded as a
function of $x^\mu$.}
\begin{eqnarray}
 \mbox{d} x^\mu \, f(x) = f(x+ \ell \, \hat{\mu}) \, \mbox{d} x^\mu
\end{eqnarray}
which shows that we are dealing with the differential calculus
of section 3 which led us to lattice (gauge) theories.\footnote{In
section 3 we considered the last relation (with $\ell \mapsto - \ell$)
on the algebra of functions on $\Rl^n$ and found that is actually lives
on a lattice. In contrast, in the present section we started with the
algebra of functions on the lattice (i.e., $\Ir^n$).} $\ell$ plays the
role of the lattice spacing.

\subsection{`Symmetric' differential calculi}
The involution introduced in section 5.1 is defined in a natural
way on the universal differential algebra. It is not consistent,
in general, with reductions of it since (\ref{e-involution})
requires that with $e_{ij} = 0$ (for some $i,j$) we must also
have $e_{ji} = 0$. A differential calculus on $\cal M$ with this
property (and also the associated graph) is called {\em symmetric}.
For such a symmetric calculus the hermitian conjugation of complex
matrices extends to matrix valued differential forms via
\begin{eqnarray}
  \phi^\dagger = \sum_{i_1 \ldots i_r} (\phi_{i_1 \ldots i_r}
                \, e_{i_1 \ldots i_r})^\dagger
              = \sum_{i_1 \ldots i_r} \phi^\dagger_{i_1 \ldots i_r}
                \, e_{i_1 \ldots i_r}^\ast  \; .
\end{eqnarray}
The condition $A^\dagger = - A$ for a connection 1-form is then
equivalent to $U^\dagger_{ij} = U_{ji}$.
\vskip.2cm
\noindent
{\em Example.} For ${\cal M} = \Ir_2 = \lbrace 0,1 \rbrace$ with the
universal differential algebra one finds
\begin{eqnarray}
   F = e_{010} \, ( U_{01} U_{10} - {\bf 1})
       + e_{101} \, ( U_{10} U_{01} - {\bf 1})
     = e_{010} \, ( U_{10}^\dagger U_{10} - {\bf 1})
       + e_{101} \, ( U_{10} U_{10}^\dagger - {\bf 1})
\end{eqnarray}
using the above condition for $A$. Then
$ S_{YM} = 2 \, \mbox{tr} \, ( U_{10}^\dagger U_{10} - {\bf 1})^2 $
which has the form of a Higgs potential (cf \cite{Conn+Lott90}).
\vskip.2cm
\noindent
{\em Remark.} Using the naive notion of dimension introduced in section
5.2, symmetric graphs are $\infty$-dimensional. This is so because
from $e_{ij}$ and $e_{ji}$ one can construct forms of arbitrarily
high grade: $e_{ijijij \ldots}$. Perhaps one should ignore such forms
in determining the dimension. Then, for example, the graph shown in
Fig.7 is quasi 1-dimensional.

\begin{minipage}{6cm}
\unitlength1.cm
\begin{picture}(5.,2.9)(-4.9,-0.4)
\thicklines
\put(0.,2.) {\circle*{0.2}}
\put(1.,1.) {\circle*{0.2}}
\put(-1.,1.) {\circle*{0.2}}
\put(1.,0.) {\circle*{0.2}}
\put(-1.,0.) {\circle*{0.2}}
\put(-1.,0.) {\line(1,0){2.}}
\put(1.,0.) {\line(0,1){1.}}
\put(1.,1.) {\line(-1,1){1.}}
\put(-1.,0.) {\line(0,1){1.}}
\put(-1.,1.) {\line(1,1){1.}}
\end{picture}
\end{minipage}
\hfill
\begin{minipage}{6cm}
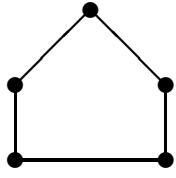
\centerline{\bf Fig.7}
\small
\noindent
A quasi-1-dimensional graph. The edges represent pairs of
antiparallel arrows.
\end{minipage}

\subsection{The symmetric lattice}
A particular example of a symmetric differential calculus is
defined as follows. We take ${\cal M} = \Ir^n$ and define a
differential calculus by
\begin{eqnarray}
 e_{ab} \neq 0 \quad \Leftrightarrow \quad
 b = a + \hat{\mu} \quad \mbox{or} \quad b = a-\hat{\mu}
 \quad \mbox{for some } \mu
\end{eqnarray}
where $\hat{\mu} = (\delta^\nu_\mu)$ and $\mu = 1, \ldots, n$.
This is represented by a hypercubic lattice graph where both arrows
are present between connected vertices. We therefore call it the
{\em symmetric lattice}. It has no distinguished directions
in contrast to the oriented lattice graph (cf section 5.4).
Furthermore, it is a lattice generalization of Connes' 2-point space.
A technical advantage over the oriented lattice is that the symmetric
lattice is compatible with the natural involution defined in section
5.1 (see also section 5.5).

As in the case of the oriented lattice we introduce
\begin{eqnarray}
           x^\mu := \ell \, \sum_a a^\mu \, e_a     \; .
\end{eqnarray}
Then
\begin{eqnarray}
 dx^\mu = \ell \, \sum_a (e_{a,a+ \hat{\mu}} - e_{a,a-\hat{\mu}}) \; .
\end{eqnarray}
Together with
\begin{eqnarray}
 \tau^\mu := \beta \, \sum_a (e_{a,a+ \hat{\mu}} + e_{a,a-\hat{\mu}})
\end{eqnarray}
(where $\beta \neq 0$ is a constant)
the $dx^\mu$ constitute a basis of $\Omega^1({\cal A})$ as a left
(or right) $\cal A$-module. For $f \in {\cal A}$ (which can be
regarded as a function of $x^\mu$) one finds the relation
\begin{eqnarray}
 df = \sum_\mu (\bar{\partial}_\mu f \, dx^\mu + {\kappa \over 2} \,
      \Delta_\mu f \, \tau^\mu )   \; .      \label{df-symlat}
\end{eqnarray}
Here $\kappa:=\ell^2/\beta$ and we have introduced the operators
\begin{eqnarray}
 \partial_{\pm \mu} f := \pm {1 \over \ell} \, \left( f(x
    \pm \ell \, \hat{\mu}) - f(x) \right) \, , \quad
 \bar{\partial}_\mu f := {1\over 2} \, (\partial_{+\mu} f +
    \partial_{-\mu} f)   \, , \quad
 \Delta_\mu f := \partial_{+\mu} \, \partial_{-\mu} f  \; .
\end{eqnarray}
For the commutation relations between functions and 1-forms we find
\begin{eqnarray}
 \lbrack dx^\mu , f(x) \rbrack & = & {\kappa \beta \over 2} \,
     \Delta_\mu f(x) \, dx^\mu + \kappa \, \bar{\partial}_\mu f(x)
     \, \tau^\mu                         \label{f,dx-comm}       \\
 \lbrack \tau^\mu , f(x) \rbrack & = &  \beta \,
     \bar{\partial}_\mu f(x) \, dx^\mu + {\kappa \beta \over 2} \,
     \Delta_\mu f(x) \, \tau^\mu \; .         \label{f,tau-comm}
\end{eqnarray}
Especially (\ref{df-symlat}) should remind us of a similar
formula in section 4.

\subsection{Graphy gravity}
In this section we anticipate a bit from a forthcoming paper
\cite{DMH94-grav}. A {\em vielbein field} on $\cal M$ is a 1-form
\begin{eqnarray}
     \vartheta = \sum_{i,j} \, e_{ij} \, \vartheta_{ij}
\end{eqnarray}
where $\vartheta_{ij} \in \Rl^n$. It is subject to local $SO(1,n-1)$
transformations,
\begin{eqnarray}
   \vartheta \mapsto G \, \vartheta
\end{eqnarray}
which implies $ \vartheta_{ij} \mapsto G(i) \, \vartheta_{ij} $.
We should therefore regard $ \vartheta_{ij} $ as sitting at the
point $i$. Using $\eta := \mbox{diag}(-1,1, \ldots,1)$ we define
\begin{eqnarray}
   (\vartheta^\dagger_{ij})_\mu := \eta_{\mu \nu} \,
   \vartheta_{ij}^\nu   \qquad (\mu = 1, \ldots, n)    \; .
\end{eqnarray}
Restricting to symmetric differential calculi (in the sense of section
5.5), we have a natural involution on $\Omega({\cal A})$. With
\begin{eqnarray}
   \vartheta^\dagger
       := \sum_{i,j} e_{ij}^\ast \, \vartheta_{ij}^\dagger
        = - \sum_{i,j} e_{ij} \, \vartheta_{ji}^\dagger
\end{eqnarray}
we find that $\vartheta \vartheta^\dagger \mapsto G \vartheta
\vartheta^\dagger G^{-1}$ under a gauge transformation.
If $F$ is the curvature of an SO(1,n-1) connection on $\cal M$, we
can now construct an analogue of the Einstein-Cartan action, namely
\begin{eqnarray}
   S_{EC} = \mbox{tr} (\vartheta \, \vartheta^\dagger , F)  \; .
\end{eqnarray}
This is indeed gauge-invariant. But what has it to do with gravity?

\section{Conclusions}
\setcounter{equation}{0}
In section 3 we have demonstrated that a simple deformation of the
algebra of differential forms on $\Rl^n$ leads us from continuum to
lattice theories. In particular, the usual action for lattice gauge
theory is obtained from the continuum Yang-Mills action via this
deformation. The `lattice differential calculus' of section 3 can
also be regarded as a differential calculus on the set of lattice
points (i.e., $\Ir^n$). In the language of section 5 it corresponds
to a hypercubic lattice graph and we have seen how everything can
be generalized to those graphs representing differential
algebras on a discrete set (which can be obtained as `reductions'
of the universal differential algebra). We have shown how the
choice of a differential algebra on a discrete set determines
a topology and assigns a dimension to it (which is actually a local
notion since it may vary from subgraph to subgraph).

In this sense, a space like ${\cal M} \times \Ir_2$ (with a manifold
$\cal M$ and the universal differential calculus on $\Ir_2$) can be
regarded as an approximation of ${\cal M} \times S^2$. The origin of a
Higgs field in the Connes \& Lott models \cite{Conn+Lott90} then
seems to be quite the same as the origin of Higgs fields in models
of dimensional reduction of gauge fields (see \cite{Mant79}, in
particular). This correspondence still has to be made more precise.

Beyond what we discussed so far, a further aspect seems to be
very promising.
One can imagine that a specific differential calculus will be
determined {\em dynamically} as a reduction of the universal
differential calculus (on a given set). This leads to a framework in
which topology change and fluctuations of dimension will naturally
occur, features which one should expect in quantum gravity.

Moreover, the differential calculus discussed in section 4 exhibited
surprising relations between bicovariant differential calculus on
quantum groups, the `proper time formalism' of quantum theory and
stochastics. It also provided us with an example of how
dynamics can be encoded in the differential calculus and then
induced on various fields.

\vskip.3cm \noindent
{\bf Acknowledgments.} I am grateful to Jerzy Lukierski
for the kind invitation and to all the organizers for the pleasant
atmosphere at the Karpacz winter school. Furthermore, I have to
thank Aristophanes Dimakis for a very fruitful and exciting
collaboration.

\end{document}